\documentclass[
aps,
reprint,
prl,
superscriptaddress,
amsmath,
amssymb,
floatfix,
longbibliography,
]{revtex4-2}

\usepackage{bm}
\usepackage{graphicx}
\usepackage{latexsym}
\usepackage{array}
\usepackage{amsfonts}
\usepackage[free-standing-units]{siunitx}
\usepackage[dvipsnames]{xcolor}
\usepackage{comment}
\usepackage{overpic}

\usepackage{mathtools}  
\usepackage{makecell}   
\usepackage{hhline}     
\usepackage{amsthm}
\usepackage{amscd}

\usepackage{xcolor}
\renewcommand\vec{\boldsymbol}

\definecolor{orange}{rgb}{1,0.5,0}
\definecolor{goodGreen's}{rgb}{0.1,0.5,0}
\definecolor{goodred}{rgb}{0.7,0,0}

\usepackage[colorlinks,urlcolor=goodGreen's,citecolor=blue,linkcolor=goodred]{hyperref}

\newcommand{\orcid}[1]{\href{https://orcid.org/#1}{\includegraphics[width=8pt]{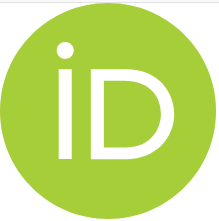}}}

\usepackage{verbatim}
\usepackage[normalem]{ulem}
\graphicspath{{./Figures_Vortices/}}

\begin{document}

\title{Abrikosov vortices in altermagnetic superconductors}

\author{A. A. Mazanik \orcid{0000-0001-6389-8653} }
\email{andrei.mazanik@csic.es}
\affiliation{Centro de Física de Materiales (CFM-MPC) Centro Mixto CSIC-UPV/EHU,
E-20018 Donostia-San Sebastián, Spain}

\author{F. S. Bergeret  \orcid{0000-0001-6007-4878}}
\affiliation{Centro de Física de Materiales (CFM-MPC) Centro Mixto CSIC-UPV/EHU, E-20018 Donostia-San Sebastián, Spain}
\affiliation{Donostia International Physics Center (DIPC), 20018 Donostia-San Sebastian, Spain}

\date{\today}

\begin{abstract}
We study the penetration of an external magnetic field into a superconductor with collinear $d$-wave altermagnetic order. We demonstrate that instead of circular Abrikosov vortices, the magnetic field generates elliptical vortices with their major axis oriented along one of the crystallographic axis, along which the altermagnetic spin splitting is maximal. 
Upon reversing the component of the magnetic field parallel to the altermagnetic N\'eel vector, the vortices reorient towards the other crystallographic axis with maximal spin splitting. We demonstrate that this effect originates from an altermagnetism-induced anisotropy of the effective mass, which is controlled by the coupling between the external magnetic field and the N\'eel vector. As a consequence, a superconducting film hosting such altermagnetic order and containing pinning defects exhibits nonreciprocal magnetization curves under reversal of the magnetic field parallel to its N\'eel vector, due to the different vortex--vortex interaction energies for the two field orientations. Our results broaden the understanding of the coexistence of altermagnetism and superconductivity, both in materials hosting these orders intrinsically or  in superconductor/altermagnet hybrid structures, and open new experimental avenues for exploring supercurrent vortices in these systems.
\end{abstract}

\maketitle

The coexistence of superconductivity and magnetism has attracted significant attention over the past decades~\cite{Bergeret_2005,Buzdin_2005,Eschrig_2015} due to their antagonistic spin orderings. With the recent identification of a new form of magnetism, altermagnetism~\cite{Smejkal2022Emerging,Smejkal2022Beyond,Tamang2025Review}, this question is being revisited in the context of superconductor/altermagnet hybrid structures~\cite{Tamang2025Review,Fukaya_2025,Junwei2025_review}. Altermagnets stand out because their electrons live in spin-split bands in momentum space while maintaining zero total magnetization. These two properties combined with superconductivity give rise to multiple phenomena \cite{Zyuzin2024,Linder2024_Memory,Belzig_2025_Thermodynamic,
Annica_BS2025_Perfect_Diode,
AnnicaBS2025_Constraints,
Kokkeler_2025,
Knolle_2025_PDW_SC_diode,Pal2025_Josephson,
bobkov2026inverse,
sachin2026altermagnetic,
schrade2026altermagnetic1,
Linder_2023_Josephson_Altermagnets,
Tanaka2024_AM_JJ,
heras2025interplay,
Justin2025_magnetoelectric,Tanaka_2025_JJ,Debnath_2026_thermoelectric,kokkeler2025nonequilibrium,vasiakin2025disorder}. Among them, altermagnetic order influences Josephson and superconducting diode effects \cite{Linder_2023_Josephson_Altermagnets,Tanaka2024_AM_JJ,Annica_BS2025_Perfect_Diode,Knolle_2025_PDW_SC_diode,heras2025interplay,Pal2025_Josephson,Tanaka_2025_JJ,sachin2026altermagnetic,schrade2026altermagnetic1}, thermoelectric effects \cite{Debnath_2026_thermoelectric}, magnetoelectric responses \cite{Zyuzin2024,Kokkeler_2025,
Justin2025_magnetoelectric,heras2025interplay}, and nonequilibrium spin-splitter effects \cite{kokkeler2025nonequilibrium}.

In Ref.~\cite{mazanik2026superconductivity}, it was shown that superconducting transport properties can serve as probes of altermagnetic order. Collinear $d$-wave altermagnetism was found to induce characteristic anisotropies in key superconducting properties, including the critical temperature, parallel critical field, and critical current. In this work, we move beyond the identification of altermagnetism in superconducting systems and investigate how it influences a fundamental topological excitation of superconductors, namely superconducting  vortices.

Supercurrent Abrikosov and Pearl vortices have been the subject of extensive research for a long time from both fundamental and technological perspectives \cite{Pearl1964_Vortices,Feigel1994,Kirtley19965_SQUID_Micro,Abrikosov2004,Wang2018Majorana,Wells2015,Kosterlitz_2016,Roditchev2015,Golod2015,Stolyarov2018,Veshchunov2016,Strunk_2022_vortices,Golod2022}. Abrikosov vortices provide the dominant contribution to the magnetization of type-II superconductors~\cite{Abrikosov2004}; the melting of a lattice of Pearl vortices in thin films leads to the Berezinskii--Kosterlitz--Thouless transition~\cite{Kosterlitz_2016}. 
In high-temperature superconductors, vortex--vortex interactions give rise to a wide variety of phases, including vortex-liquid and vortex-glass states~\cite{Feigel1994}. 
Beyond their fundamental importance, vortices can serve as information carriers~\cite{Golod2015,Golod2022}, enable supercurrent non-reciprocity~\cite{Golod2022,hou2023ubiquitous}, and can host Majorana states in their cores in topological superconductors~\cite{Wang2018Majorana}.

In this Letter, we show how superconducting vortices are modified by the coexistence of superconductivity and collinear $d$-wave altermagnetic order.  We first demonstrate that vortices acquire an elliptical shape, with the ellipticity controlled by the relative orientation of the external magnetic field and the N\'eel vector.   We further show that field-controlled vortex ellipticity makes the vortex--vortex interaction sensitive to the orientation of the magnetic field relative to the N\'eel vector, resulting in different interaction energies for parallel and antiparallel fields. As a consequence, when vortex motion is constrained by geometry or pinning, this anisotropy leads to nonreciprocal magnetization curves under field reversal, providing an experimentally accessible signature of altermagnetic vortex ellipticity.


{\it System and model.}
We consider a superconductor hosting a collinear $d$-wave altermagnetic order, either intrinsic or induced via magnetic proximity to an altermagnetic insulator (AMI) \cite{vasiakin2025disorder,Belzig_2025_Thermodynamic,heras2025interplay,bobkov2026inverse}.
The superconductor is assumed to be infinite in the $\hat{\vec{x}}$ and $\hat{\vec{y}}$ directions and it is subjected to an external magnetic field directed along the $\hat{\vec{z}}$ axis. The magnetic field penetrates the sample in the form of Abrikosov vortices, implying that the film thickness $d_S$ along the $\hat{\vec{z}}$ axis is sufficiently large. In order to describe the response of the system to the applied field, we employ the Ginzburg-Landau free energy functional over the superconducting order parameter, $F[\Psi]$, recently derived for such systems \cite{Zyuzin2024,heras2025interplay} that reads as:
\begin{align}
    &F[\Psi] = \int dV\ \mathcal{F} = \int dV\,\left\{
 a \vert \Psi \vert^2
+ \frac{b}{2} \vert \Psi \vert^4 + \frac{B^2}{8\pi}    \right. \nonumber\\
 &\ \ \left.
 + \frac{1}{2m^*} \left( D_k \Psi\, D_k^\star \Psi^\star+B_a N_a  K_{jk}  D_j  \Psi\, D_k^\star \Psi^\star \right) \right\}, \label{eq:GL_Free_Energy}
\end{align}
where the integration is performed over the superconductor. Here, $a = \alpha (T - T_{c0}) < 0$, $b$, and $m^\star$ are the GL coefficients renormalized by the altermagnetic order \cite{heras2025interplay}. $T$ and $T_{c0}$ are the system temperature and the field-free critical temperature, respectively, and $\alpha >0$. The magnetic field is given by $\vec{B}=\nabla\times\vec{A}$, where $\vec{A}$ is the vector potential. The gauge-covariant derivative reads $D_k=-i\nabla_k-\frac{e^*}{c}A_k$, where $e^*=2e$ is the charge of a Cooper pair and $c$ is the speed of light.

The collinear $d$-wave altermagnetic order is characterized by the tensor $N_a K_{jk}$, where $\vec{N}$ is the unit N\'eel vector defining the spin-quantization axis of band electrons, and $K_{jk}$ determines the directions of spin splitting. By definition, $\vec{N}$ is odd under time reversal, whereas $K_{jk}$ is time-reversal even and symmetric in its indices. The N\'eel vector can point in an arbitrary direction; for simplicity, we choose it along the $\hat{\vec{z}}$ axis, $\vec{N}=(0,0,1)$. Without loss of generality, we take the tensor $K_{jk}$ in such a form that it has only two non-vanishing components, $K_{xx}=-K_{yy}=K$ corresponding to maximal spin splitting along the  $\hat{\vec{x}}$ and $\hat{\vec{y}}$ axes in the momentum space~\cite{heras2025interplay}. Physically, the last term in Eq.~(\ref{eq:GL_Free_Energy}) can be interpreted as a renormalization of the effective mass, resulting in an anisotropy controlled by the magnetic field.

After varying the GL functional Eq.~\eqref{eq:GL_Free_Energy} with respect to $\Psi^\star$ and $\vec{A}$, we obtain the  GL equations for an altermagnetic superconductor:
\begin{subequations} \label{eq:main}
\begin{align}
    &a \Psi + b \vert \Psi \vert^2 \Psi -\frac{1}{2m^*} \left( D^2_k \Psi + D_k \left\{ B_a N_a K_{jk}  D_j \Psi  \right\} \right)  = 0, \label{eq:Psi_Eq}\\
    & \epsilon_{kjm}\nabla_j (B_{m} - 4\pi \mathcal{M}_m)  = \frac{4\pi }{c} j_{Sk}, \label{eq:rot_H} \\
    & j_{Sk} = \frac{e^*}{2m^* }\left\{   \delta_{jk} + B_a N_a K_{jk} \right\} \left( \Psi D^\star_j \Psi^\star + \Psi^\star D_j \Psi \right), \label{eq:j_S} \\
    & \mathcal{M}_m = - \frac{\partial }{\partial B_m }\left(\mathcal{F} - \frac{B^2}{8\pi}\right) = -\frac{1}{2m^*} N_m K_{jk}D_j \Psi D^\star_k \Psi^\star.  \label{eq:M_def}
\end{align}
\end{subequations}
Here, $\vec{j}_S$ denotes a supercurrent, and $\vec{\mathcal{M}}$ represents the magnetic moment associated with the altermagnetism which is generated by an inhomogeneity of the phase or modulus of the superconducting order parameter, $\Psi = \vert \Psi \vert e^{i\varphi}$ \cite{Zyuzin2024,Kokkeler_2025,
Justin2025_magnetoelectric,heras2025interplay}. The system of Eqs.~\eqref{eq:main}, together with the corresponding boundary conditions stating the absence of supercurrents through the system edges, $\left. \vec{n}\cdot \vec{j}_S \right\vert_{z=\pm d_S/2} = 0$ with $\vec{n} = (0,0,\pm 1)$, describe the response of the superconducting film to the applied magnetic field.

{\it Results.} We start investigating supercurrent vortices in collinear $d$-wave altermagnetic superconductors from the case of an isolated Abrikosov vortex generated inside a thick film, $d_S \gg \xi,\lambda_L$, where $\xi$ and $\lambda_L$ are the GL coherence length and the London penetration depth, respectively. This vortex is characterized by the field distribution $\vec{h} = \vec{B} - 4\pi \vec{\mathcal{M}}= (0,0,h)$. Within the London approximation, defined by retaining terms up to the first order in the superfluid velocity $\xi\left\vert \nabla\varphi - e^* \vec{A}/c\right\vert \propto \xi/\lambda_L \ll 1$, we obtain the following equation for $h(x,y)$, valid far away from the vortex core located at $\vec{r}_a$:
\begin{equation}
    h -  \lambda^2_L \left[ \partial_x \left(\frac{\partial_x h}{1 - K h}\right) + \partial_y\left(\frac{\partial_y h}{1 + K h}\right) \right] =  \Phi_0 q_a  \delta(\vec{r} - \vec{r}_a), \label{eq:H_eq}
\end{equation}
where $\lambda^2_L = \frac{m^* c^2}{4\pi {e^*}^2 \vert \Psi\vert^2}$, $\Phi_0 = \frac{\pi c}{e}$ is the flux quantum, and $q_a = \pm 1$ characterizes the winding number of the superconducting phase $\varphi$ along the vortex core. $q$ is determined by the direction of the external magnetic field $\vec{H}$: $q_a = \vec{H}\cdot\hat{\vec{z}}/H$. When multiple vortices penetrate the sample, then in the right hand side of Eq.~\eqref{eq:H_eq} one has to write the sum over vortices, $\sum_a q_a \delta(\vec{r} - \vec{r}_a)$.

\begin{figure}[thbp]
    \centering
    \includegraphics[height=3.6cm]{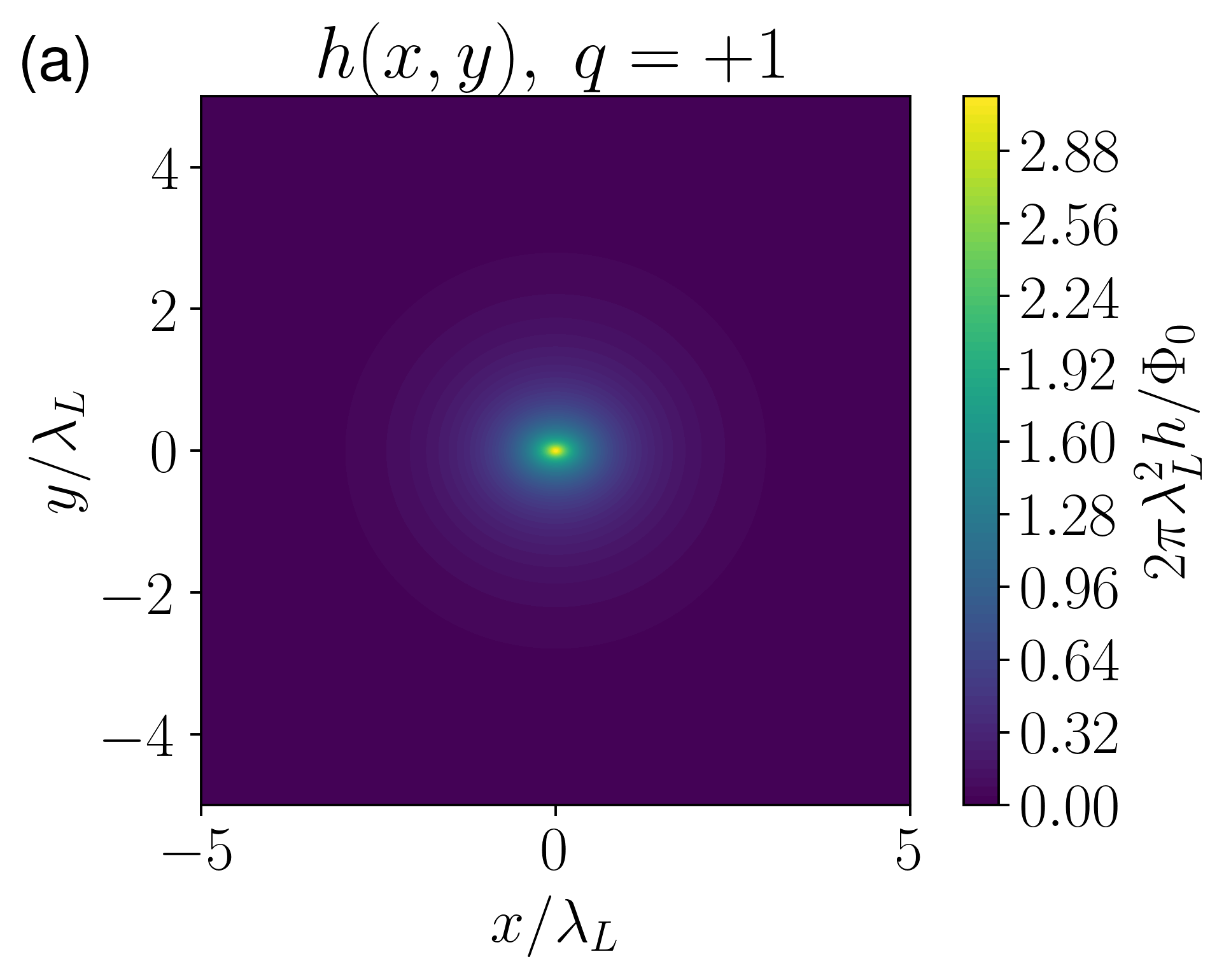}
    \includegraphics[height=3.6cm]{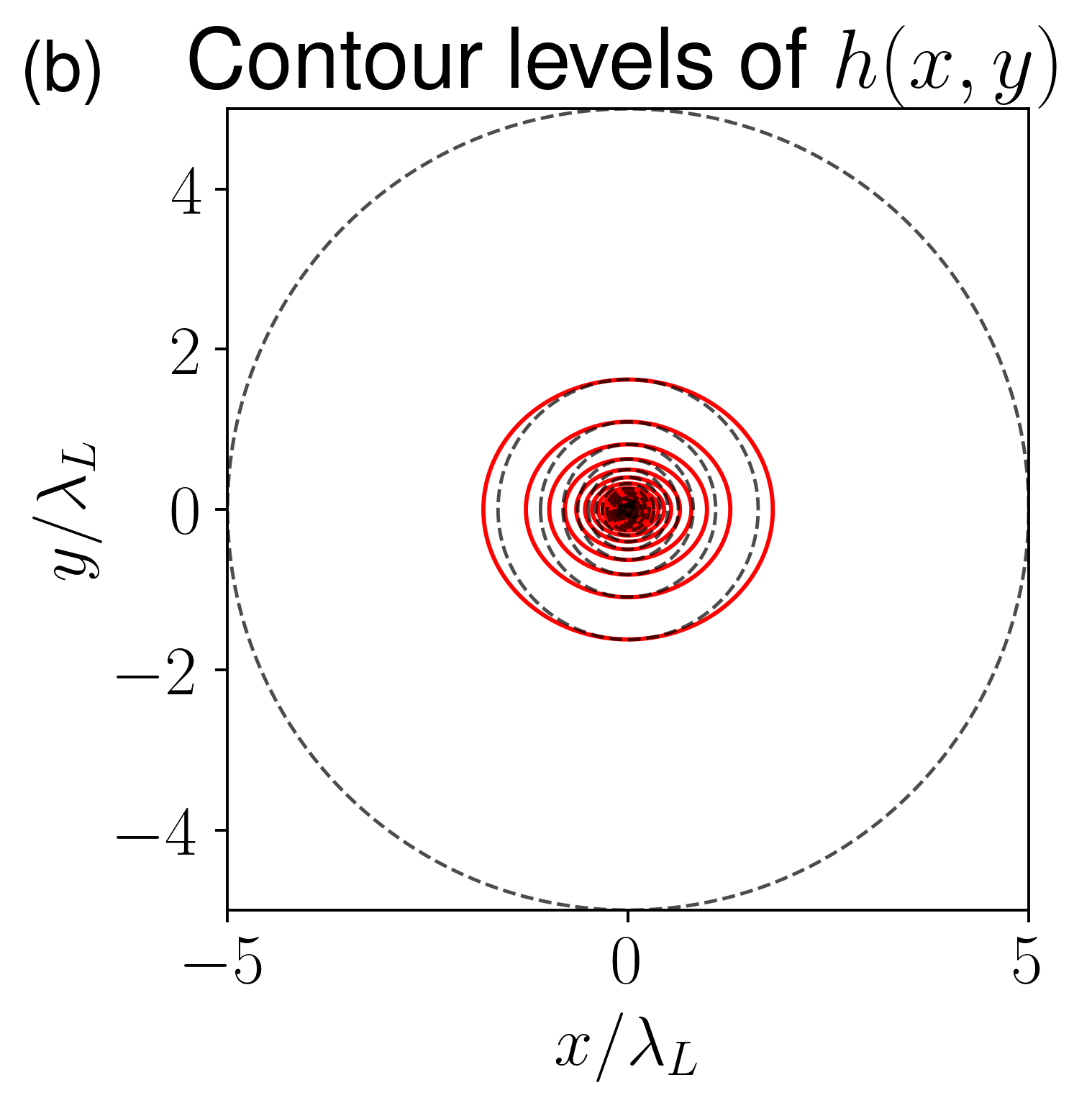}
    \includegraphics[height=3.6cm]{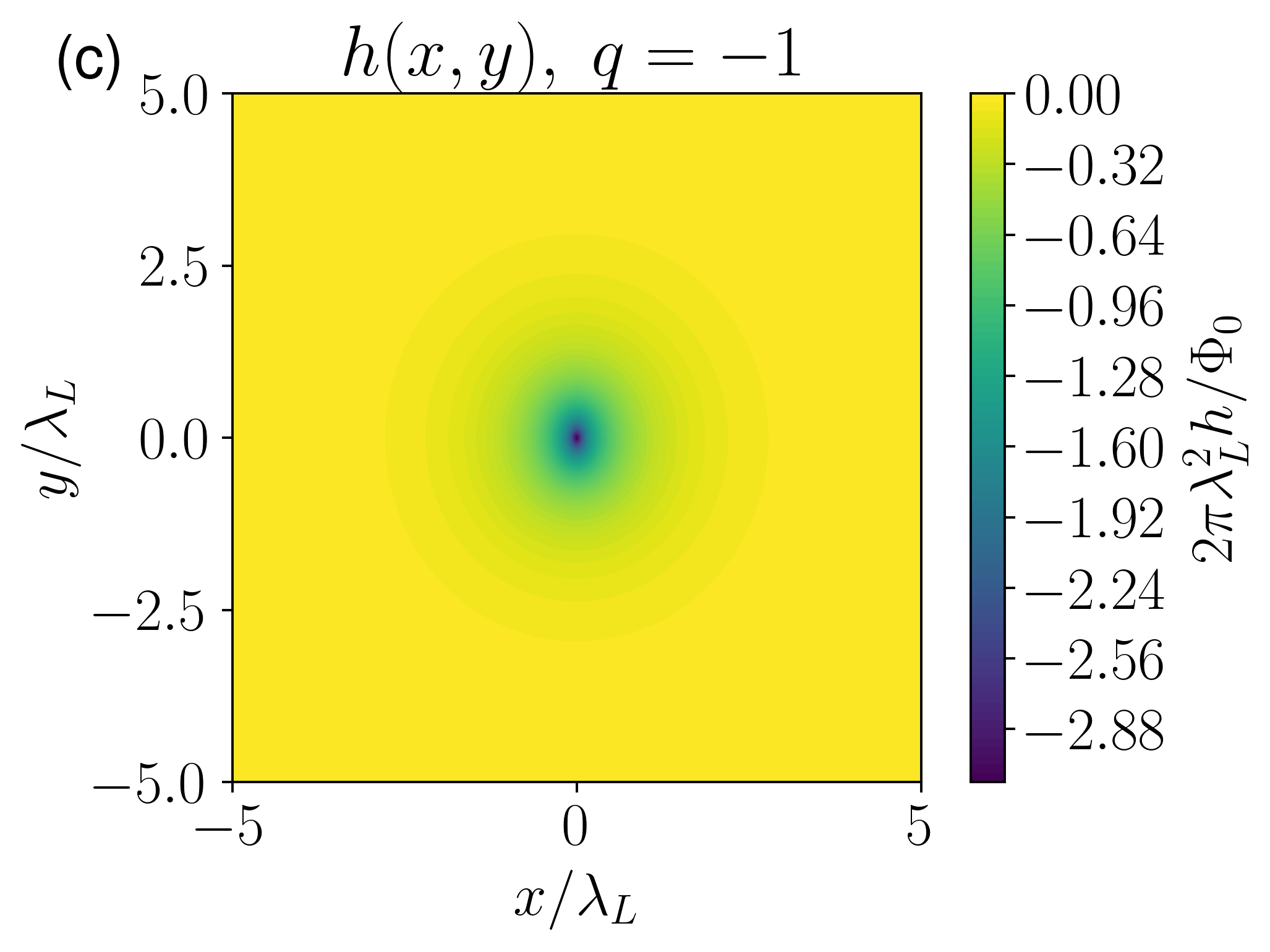}
    \includegraphics[height=3.6cm]{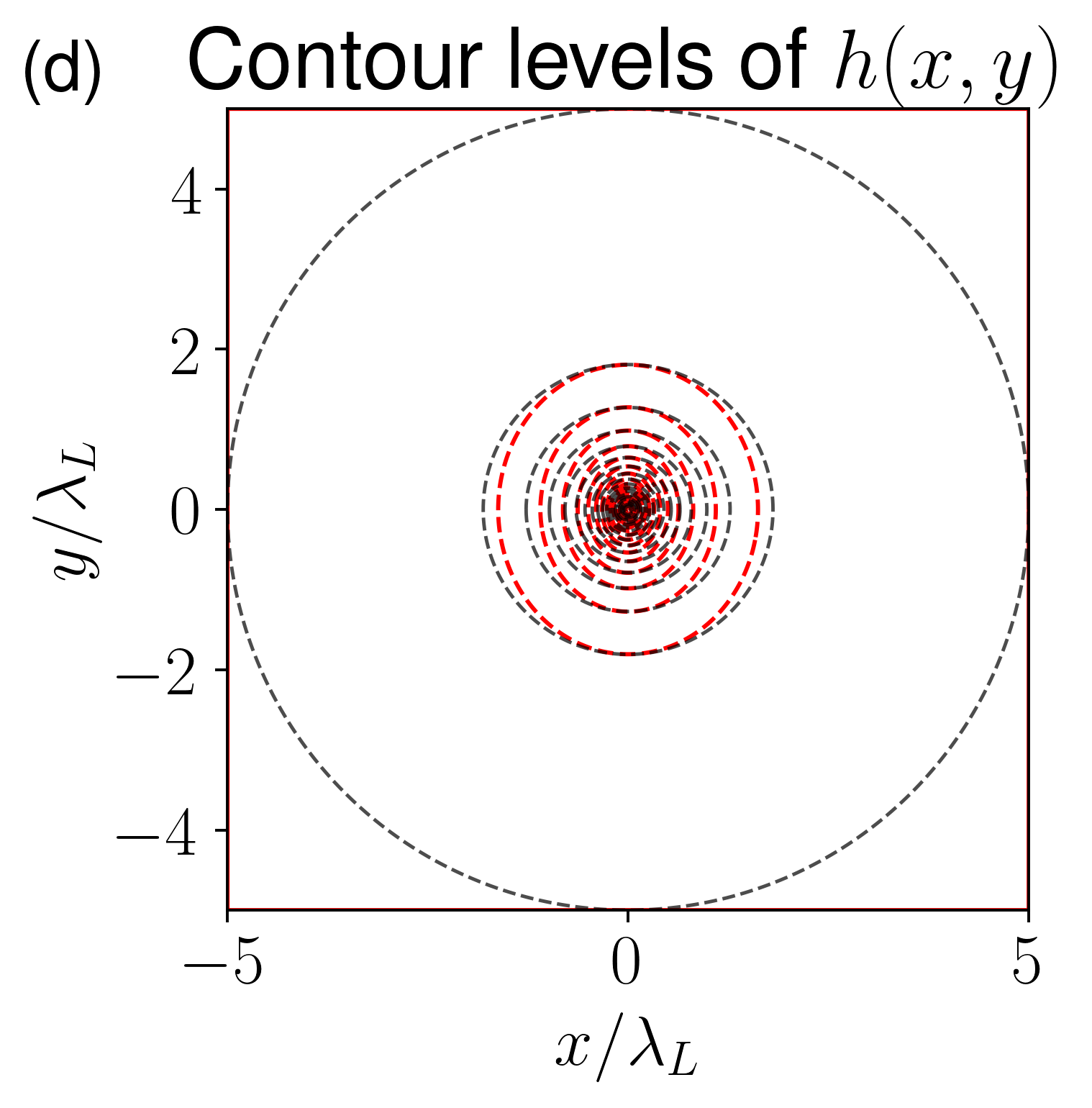}
    \caption{Field configurations of a vortex located at $\vec{r}_a = 0$ with $q=+1$ [(a),(b)] and an antivortex with $q=-1$ [(c),(d)], obtained from Eq.~\eqref{eq:H_eq} by  imposing  the boundary condition $h(|\vec{r}| \to \infty) \to 0$. In (b) and (d), contour levels of the magnetic field $h(x,y)$ are shown in red, while black dashed lines indicate reference circles highlighting the elliptical shape of the $h(x,y)$ profiles.  The altermagnetic parameter is set to $K\Phi_0/(2\pi \lambda_L^2)=0.2$.}
    \label{fig:H_xy}
\end{figure}

The Eq.~(\ref{eq:H_eq}) is symmetric with respect to the $x\to -x$ and $y \to -y$ operations when $\vec{r}_a = 0$. This fact implies that its solutions, $h(x,y)$, obey the same symmetries that result in an elliptical form of vortices ($q = +1$) and antivortices ($q = -1$). The numerical solutions of Eq.~(\ref{eq:H_eq}) presenting such elliptical vortices are shown in Figs.~\ref{fig:H_xy}(a)--(d). We see that the vortex is oriented along the $\hat{\vec{x}}$ axis, while the antivortex  prefers the orientation along the $\hat{\vec{y}}$ axis.  We emphasize that this property distinguishes vortices in collinear $d$-wave altermagnetic superconductors from those in conventional anisotropic superconductors~\cite{Larkin1992,Walker_1959_BCS_overlapping_bands,Bhandari_1972_SC_Magnetoresistance}, where the effective mass tensor is independent of the external magnetic field, and therefore the field-controlled ellipticity of vortices is absent in them.

We extend our analysis to the structure of vortex cores in collinear $d$-wave altermagnetic superconductors. In conventional type-II superconductors, vortices form narrow threadlike regions with a characteristic diameter determined by the GL coherence length $\xi = 1/\sqrt{2m^* \vert a \vert}$. Within these vortices, $|\Psi|^2 \ll 1$, while the magnetic field is approximately constant, $h \approx \mathrm{const} \propto \Phi_0/(2\pi \lambda_L^2)$~\cite{tinkham2004introduction,schmidt2013physics}. Guided by these properties, we neglect the nonlinear term $|\Psi|^2\Psi$ and approximate $B_a N_a \approx h_a N_a = h_0$ as constant in Eq.~\eqref{eq:Psi_Eq} near a vortex core. 
Rescaling the coordinates as $X = x/[\xi\sqrt{1+Kh_0}]$ and $Y = y/[\xi\sqrt{1-Kh_0}]$, and retaining the leading terms in the resulting equation~\cite{tinkham2004introduction}, we obtain the condensate density profile
\begin{equation} \label{eq:V_profile}
    \vert \Psi(x,y)\vert ^2 \propto \frac{(x/\xi)^2}{1+Kh_0} + \frac{(y/\xi)^2}{1-Kh_0}.
\end{equation}
Here, $h_0 \propto \Phi_0/(2\pi \lambda_L^2)$ is the  constant magnetic field inside the vortex core, which is of the order of the first critical field of the altermagnetic superconductor. When the direction of the applied magnetic field is reversed, $h_0$ changes its sign in Eq.~\eqref{eq:V_profile} accordingly to the magnetic field profiles shown in Figs.~\ref{fig:H_xy}(a)--(d).

The ellipticity of Abrikosov vortices controlled by the direction of the external magnetic field has direct consequences for observable quantities. For example, one may measure local magnetic fields \cite{Kirtley19965_SQUID_Micro,Wells2015}, or STM may be used to examine the spatial structure of vortex cores predicted by Eq.~\eqref{eq:V_profile} \cite{Stolyarov2018,Roditchev2015}. Other probes of the vortex structure include film magnetization $M(H)$~\cite{schmidt2013physics,tinkham2004introduction} and vortex inductance~\cite{Strunk_2022_vortices}.

\begin{figure}[thbp]
    \centering
    \includegraphics[width=0.95\linewidth]{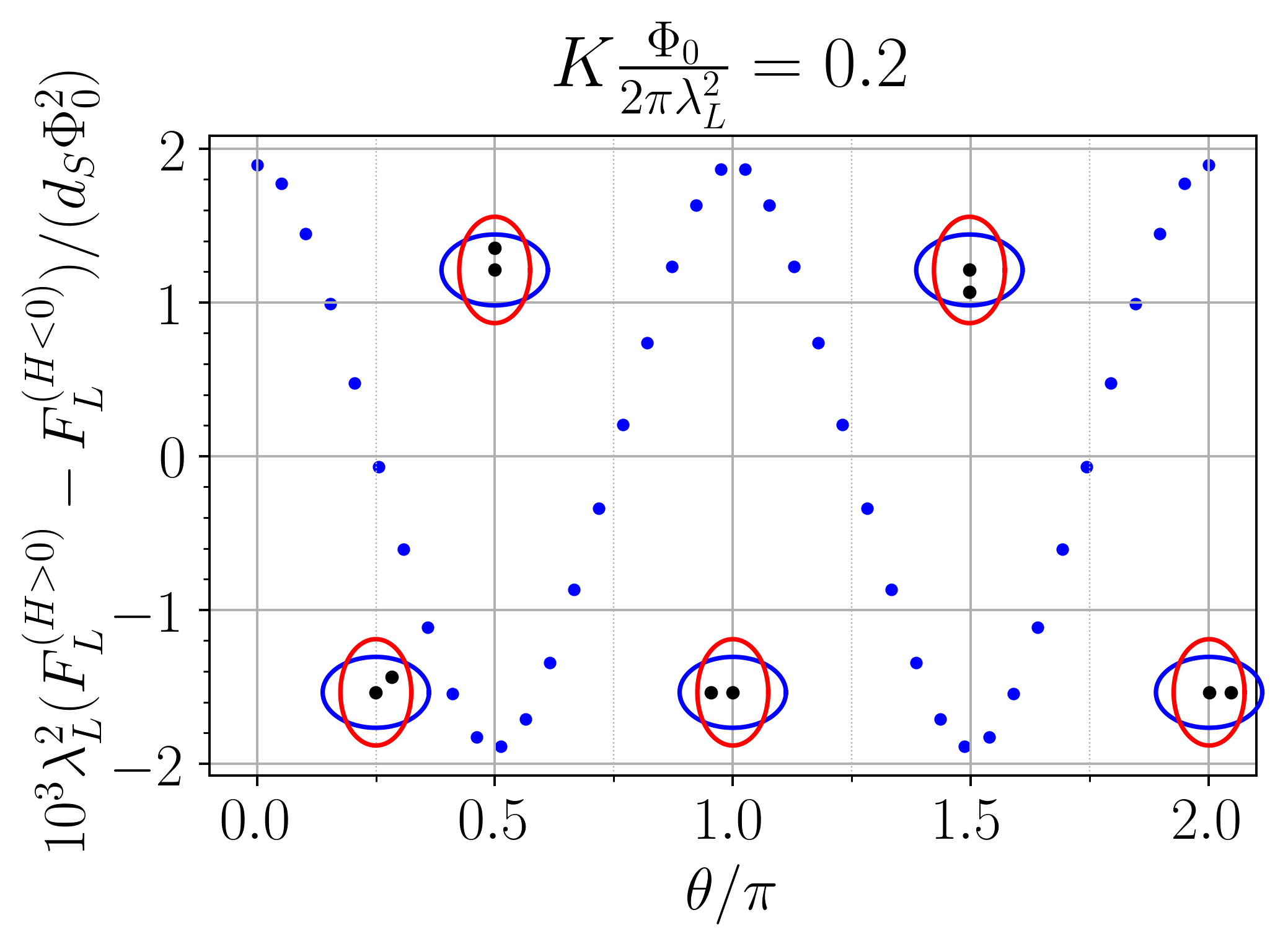}
    \caption{ { The free-energy difference between the film states with a pair of vortices and a pair of antivortices,} $F^{(H>0)}_L - F^{(H<0)}_L$.  One of the vortices is pinned at $\vec{r}_0=(0,0)$ and the other is rotated around it, $\vec{r}_1=(d\cos\theta,d\sin\theta)$ with $d=\lambda_L$. 
    The difference $F_L^{(H>0)} - F_L^{(H<0)}$ vanishes at $\theta=(1+2n)\pi/4$, where $n=0,\pm1,\dots$, and reaches its extrema at $\theta = n\pi/2$. Insets show the relative orientation of the two vortices [or antivortices] with respect to the altermagnetic splitting configuration. }
    \label{fig:VV_potential}
\end{figure}

In type-II superconducting films, the dominant contribution to magnetization $M(H)$ arises from supercurrent vortices and their spatial configuration~\cite{schmidt2013physics,tinkham2004introduction}. 
In extended films free of pinning defects, this configuration is governed by vortex--vortex interactions. To analyze the interaction potential in our altermagnetic superconductors, we derive from Eqs.~\eqref{eq:main}(a)--(d) the London free-energy functional,
\begin{equation} \label{eq:Free_London}
    \frac{8\pi F_L}{d_S}=  \int_{\mathrm{film}} dx\,dy \left\{h^2 + \lambda_L^2 \left[\frac{(\partial_x h)^2}{1 - Kh} + \frac{(\partial_y h)^2}{1 + Kh} \right]\right\}.
\end{equation}
Here, $h(x,y)$ is the solution of Eq.~\eqref{eq:H_eq} with the source term $\Phi_0 \sum_a q_a \delta(\vec{r}-\vec{r}_a)$, and the integration is carried out along the film. The vortex--vortex interaction is governed by the magnetic field distribution between the vortex cores. Let us first focus on two vortices. When their ellipticities enhance the magnetic field in the inter-core region, the interaction energy increases relative to that of circular vortices. This occurs when the axis connecting the two vortices is at $\theta = 0, \pi$ for the configuration shown in the inset of Fig.~\ref{fig:VV_potential}.
In contrast, when the ellipticities reduce the magnetic field between the cores [$\theta = \pi/2, 3\pi/2$], the interaction energy is correspondingly lowered. 

Interestingly, the interaction becomes nonreciprocal when the direction of the external field is inverted. To demonstrate this, we use Eq.~\eqref{eq:Free_London} to compute the energy difference between a film state containing two vortices and a state with two antivortices (opposite external field) at the same positions, $F^{(H>0)}_L - F^{(H<0)}_L$, as a function of their relative orientation at a fixed distance between vortex cores. This calculation is shown in Fig.~\ref{fig:VV_potential}. The interaction energies for vortices and antivortices coincide only when the pairs are aligned along the nodes of the altermagnet, $\theta = (1+2n)\pi/4$, $n=0,\pm1,\dots$, in the configuration of Fig.~\ref{fig:VV_potential}. In contrast, for any other value of $\theta$ the difference between the interaction energies is finite. In particular, 
when pairs are aligned along axes with maximal spin splitting, this  difference for  $H>0$ and $H<0$ shows extrema. 


We note that the properties of the vortex--vortex potential inferred from Fig.~\ref{fig:VV_potential} also lead to the fact that the vortex--vortex potential is invariant under a simultaneous transformation of vortices into antivortices and a $\pi/2$ rotation of the vortex cores. This symmetry is also present in Eq.~\eqref{eq:H_eq}. Consequently, in films without constraints on vortex motion, the response of the superconductor is reciprocal with respect to switching the field direction, \textit{i.e.}, the magnetization curves remain symmetric under field reversal, since the vortex lattice is simply rotated by $\pi/2$ under $H \to -H$. In contrast, this symmetry may be broken when vortex motion is restricted by geometry, e.g., in island geometries \cite{Roditchev2015}, or by pinning centers.

\begin{figure}[thbp]
    \centering
    \includegraphics[width=0.9\linewidth]{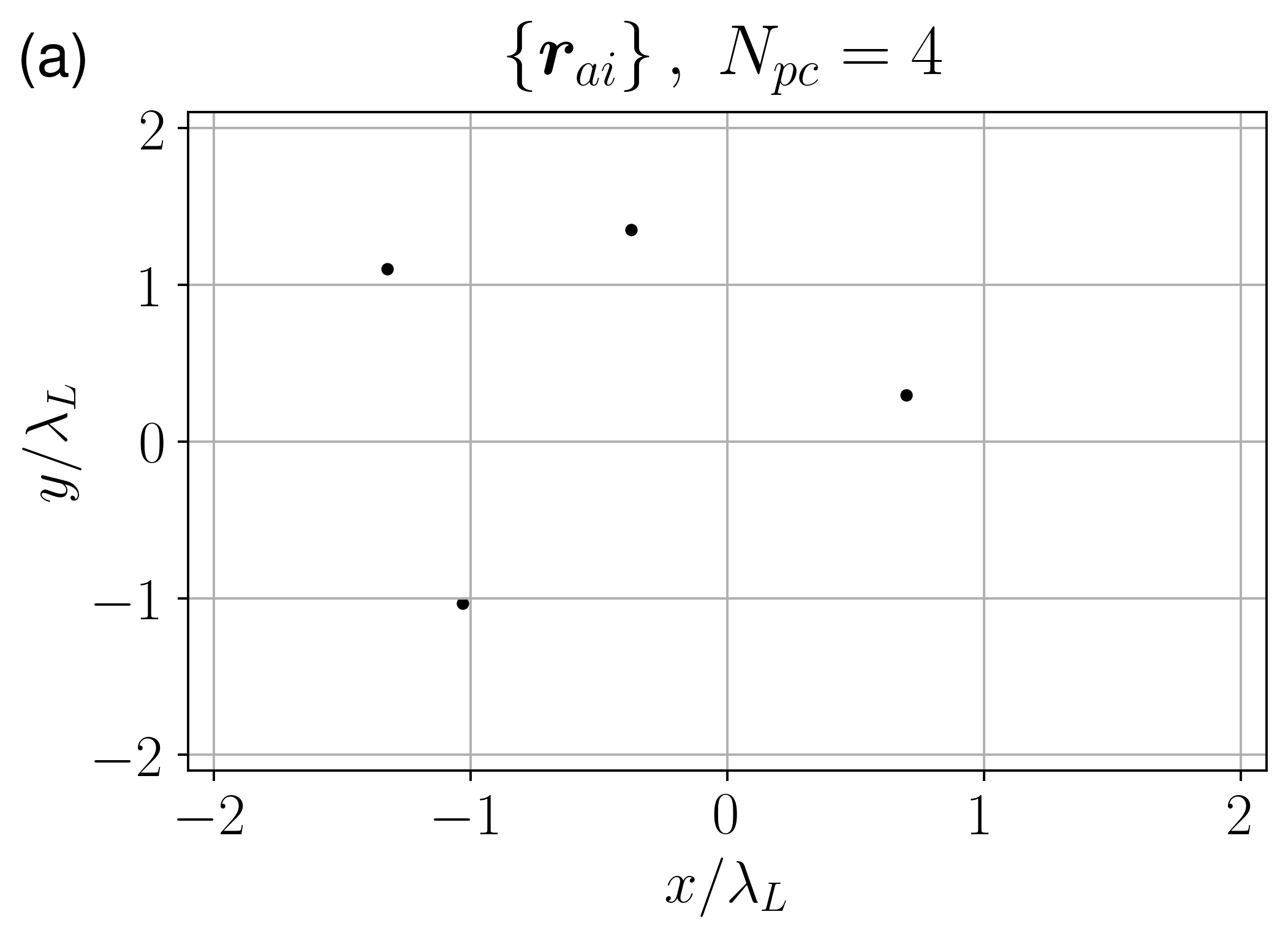}
    \includegraphics[width=0.9\linewidth]{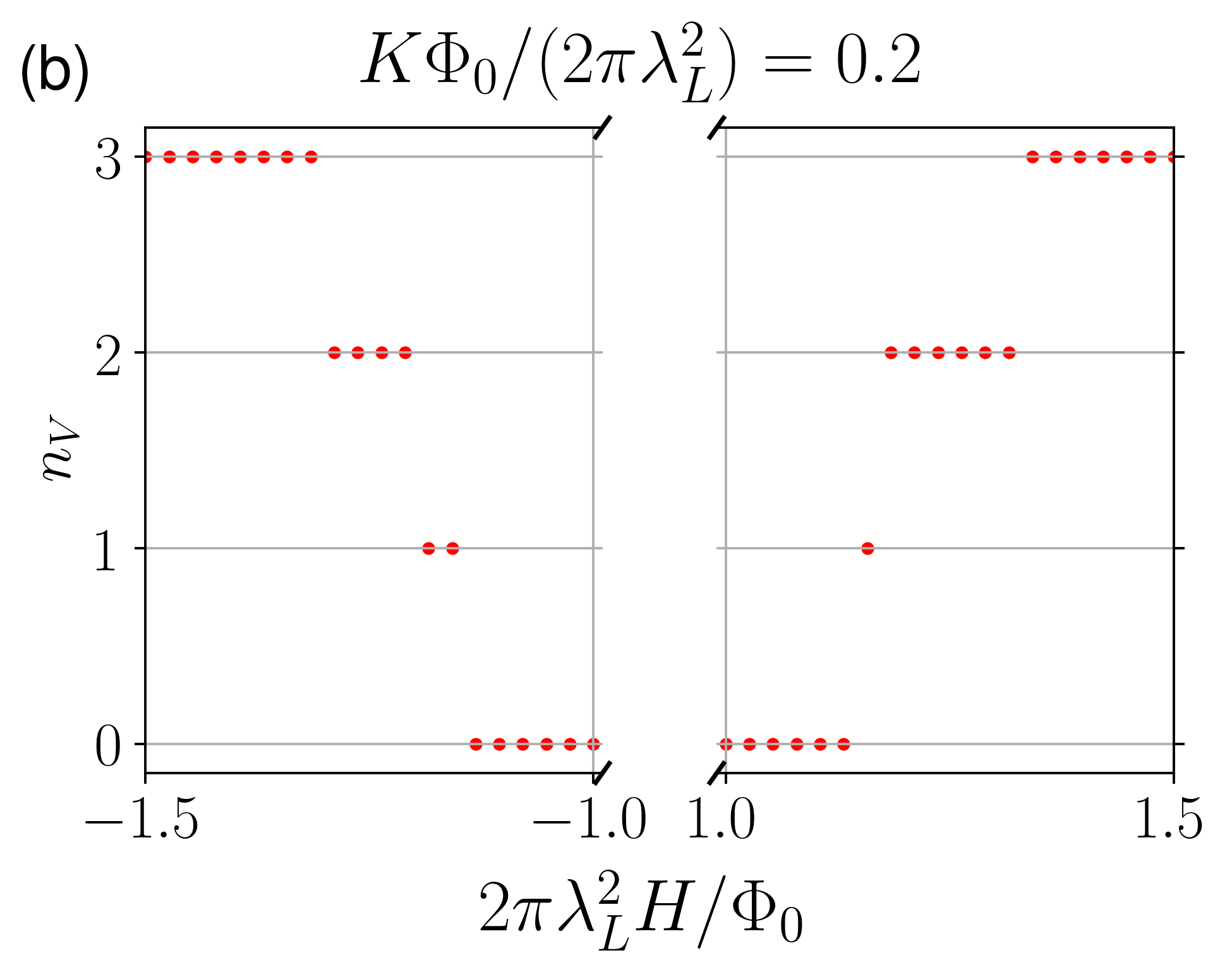}
    \includegraphics[width=0.9\linewidth]{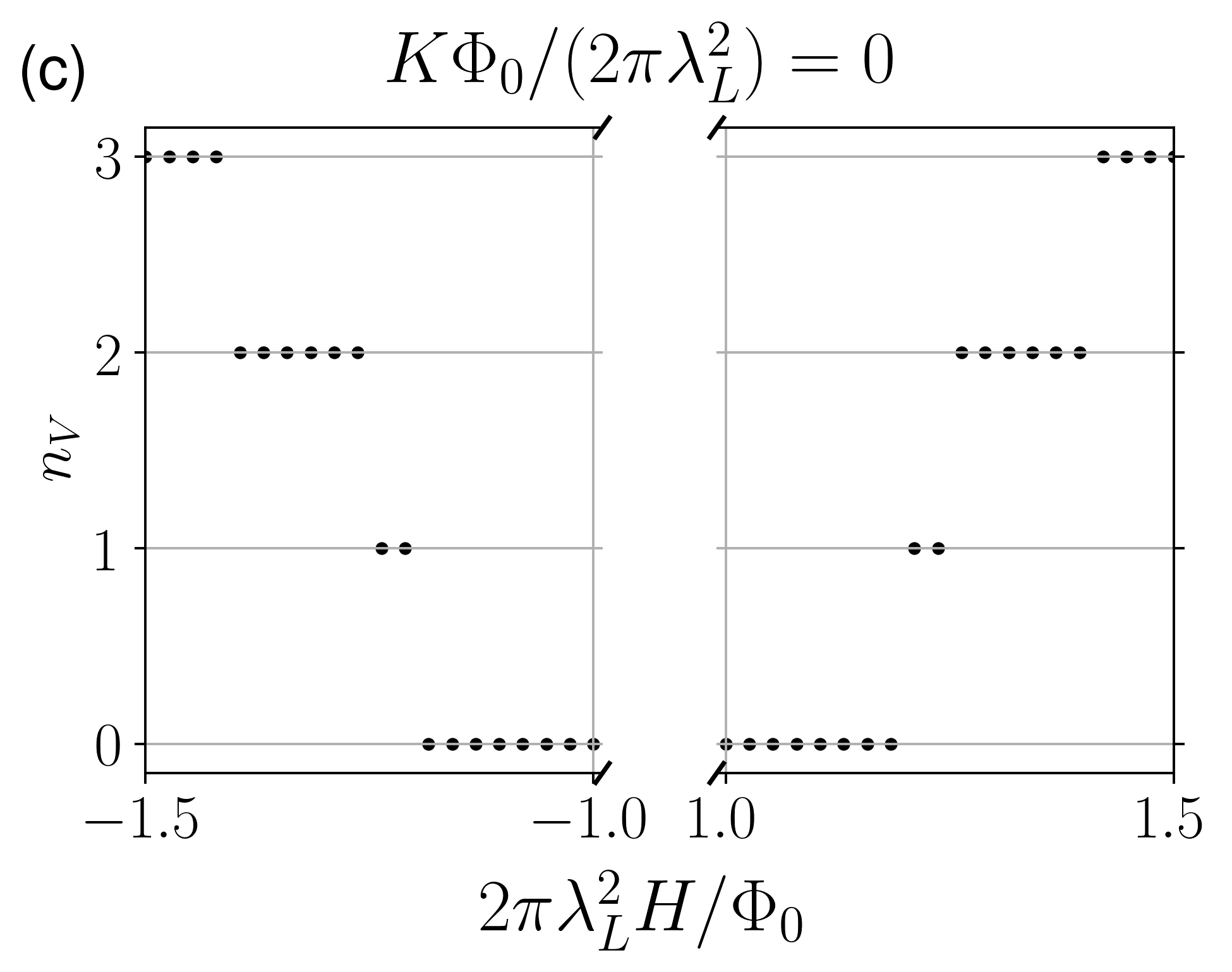}
    \caption{(a): Spatial map of the pinning centers used in the calculations, $N_{pc}=4$. (b),(c): Field dependencies of the vortex number $n_V(H)$ for a superconducting film with (b) and without (c) the altermagnetic order according to the energy functional given by Eqs.~\eqref{eq:Gibbs} and \eqref{eq:Free_London}. For $\vert H \vert <H_{c1}\propto \Phi_0/(2\pi \lambda_L^2)$, vortices do not enter the sample.  }
    \label{fig:n_V_H}
\end{figure}

To understand how pinning centers or geometrical restrictions can influence  the symmetry of the magnetization curves of superconducting films with the altermagnetic order, we
consider a finite film described by the functional Eq.~\eqref{eq:GL_Free_Energy}, in which only a finite number of pinning centers $N_{pc}$ exist, which are located at $\left\{ \vec{r}_{ai}\right\}$, $i \in 1...N_{pc}$. $\left\{ \vec{r}_{ai}\right\}$ are taken randomly within a fixed square $3\lambda_L \times 3\lambda_L$.  If a vortex enters the system, then we assume that it is pinned by one of the existing pinning centers. For a given external magnetic field $H$ and the configuration of pinning centers $\left\{ \vec{r}_{ai}\right\}$, we numerically compute  the Gibbs potential for each possible vortex or antivortex distribution along these centers and pick the vortex  configuration that corresponds to the minimum of the Gibbs potential. Finally, we obtain the dependence of the number of  existing vortices on the external field, $n_V(H)$, for the given disorder configuration. The Gibbs potential, which we use for this, reads as
\begin{equation} \label{eq:Gibbs}
    \frac{ 8 \pi  G(n_V, H)}{d_S } = \frac{8\pi F_L}{d_S}  - 2n_V \Phi_0 H.
\end{equation}
Here, $F_L$ refers to the free energy given by Eq.~\eqref{eq:Free_London} with the corresponding vortex configuration.

The typical results of the $n_V(H)$ calculation based on Eq.~\eqref{eq:Gibbs} for fully pinned vortices are shown in Figs.~\ref{fig:n_V_H}(a)--(c). We find that, in both altermagnetic superconductors and conventional superconductors without altermagnetism, the first critical fields for vortices and antivortices coincide, $H_{c1}^{(q=+1)} = H_{c1}^{(q=-1)}$, reflecting the equal energies of isolated vortices and antivortices. However, for $H > H_{c1}$, the symmetry $n_V(H) = n_V(-H)$ is broken for $K \neq 0$ due to vortex--vortex interactions, which differ for vortices and antivortices, as discussed above. As a consequence, the  magnetization curve becomes nonreciprocal under field reversal $M(H) \neq M(-H)$, for $K \neq 0$ and fields slightly above $H_{c1}$. In contrast, a conventional superconducting sample with identical pinning centers exhibits symmetric behavior $n_V(H)=n_V(-H)$ [Fig.~\ref{fig:n_V_H}(c)], and thus its magnetization obeys $M(H)=M(-H)$.

Finally, we note that, despite the asymmetry $n_V(H) \neq n_V(-H)$, the second critical field of the altermagnetic superconductor remains independent of the direction of the applied field. To demonstrate this, we linearize the GL equation~\eqref{eq:Psi_Eq} with respect to $\Psi$ and determine its lowest eigenvalue, corresponding to the nucleation of superconductivity in a uniform magnetic field $\vec{h}=H^{(AM)}_{c2}\hat{\vec{z}}$~\cite{tinkham2004introduction,schmidt2013physics}. 
This procedure yields
\begin{equation}
H^{(AM)}_{c2} = H_{c2}\left(1 + \frac{K^2 H_{c2}^2}{2} \right),
\end{equation}
where $H_{c2} = \Phi_0/(2\pi \xi^2)$ is the upper critical field in the absence of altermagnetism. 

{\it Conclusions and discussion. }
In summary, we have investigated the formation of supercurrent vortices in collinear $d$-wave altermagnetic superconductors. We have found that altermagnetic order induces a pronounced ellipticity of the vortices, controlled by the direction of the applied magnetic field relative to the N\'eel vector and the crystallographic axes with maximal spin splitting. This field-induced ellipticity can be probed via local magnetic field measurements, through density-of-states measurements (e.g., STM), or through its impact on magnetization curves, which become nonreciprocal under reversal of the corresponding magnetic field component when the vortex movement is restricted by pinning defects or by a sample geometry. It is plausible to expect that similar physics extends to the vortex inductance \cite{Strunk_2022_vortices}, and  to thin superconducting films, affecting Pearl vortices~\cite{Pearl1964_Vortices} and potentially leading to an asymmetric Berezinskii--Kosterlitz--Thouless transition~\cite{Kosterlitz_2016}. All these effects can arise in hypothetical altermagnetic superconductors, as well as in bilayers composed of conventional BCS superconducting films (e.g., Al or Nb) placed on top of altermagnetic insulators~\cite{Smejkal2022Emerging}, or antiferromagnets with surface-emergent altermagnetic order recently predicted in Ref.~\cite{lange2026emergent}. 


{\it Acknowledgments.}
We thank Tim Kokkeler, Rodrigo de las Heras, Ilya Tokatly, and Vitaly Golovach for useful discussions. F.~S.~B. thanks financial support from the the Spanish MCIN/AEI/10.13039/501100011033 
through the grants PID2023-148225NB-C31, and TED2021-130292B-C41. A. A. M. and F. S. B. also thank financial support from the
European Union’s Horizon Europe research and innovation program under grant agreement No. 101130224 (JOSEPHINE). 

\bibliographystyle{apsrev4-2}
\bibliography{refs}

\end{document}